# Tunable Magnetic Domain Walls for Therapeutic Neuromodulation at Cellular Level: Stimulating Neurons Through Magnetic Nanowires

Diqing Su[1, #], Kai Wu[2, #], Renata Saha[2], and Jian-Ping Wang[2, *]

[1]Department of Chemical Engineering and Material Science, University of Minnesota, Minneapolis, Minnesota 55455, USA

[2]Department of Electrical and Computer Engineering, University of Minnesota, Minneapolis, Minnesota 55455, USA

[#]These authors contributed equally to this work

*Corresponding author E-mail: jpwang@umn.edu

**Abstract:** Cellular-level neuron stimulation has attracted much attention in the areas of prevention, diagnosis and treatment of neurological disorders. Herein, we propose a spintronic neurostimulator based on the domain wall movement inside stationary magnetic nanowires driven by the spin transfer torque. The electromotive forces generated by the domain wall motion can serve as highly localized stimulation signals for neuron cells. Our simulation results show that the induced electric field from the domain wall motion in permalloy nanowires can reach up to $14\ V/m$, which is well above the reported threshold stimulation signal for clinical applications. The proposed device operates on a current range of several $\mu A$ which is $10^3$ times lower compared to magnetic stimulation by microcoils. The duration and amplitude of the stimulating signal can be controlled by adjusting the applied current density, the geometry of the nanowire, and the magnetic properties of the nanowire material.

## 1. Introduction

In 2013, the National Institutes of Health (NIH), the Defense Advanced Research Projects Agency (DARPA), and the National Science Foundation (NSF) launched a project named BRAIN Initiative [1,2] to accomplish the prevention, diagnosis and treatment of brain disorders such as Alzheimer's disease, attention deficit hyperactivity disorder (ADHD), Parkinson's disease, migraine, and traumatic brain injury (TBI). The primary challenge in this process is the lack of understanding of the pathogenesis, which makes it necessary to investigate the interactions within the brain from cellular level to the complex neural circuits through brain stimulation. Since the pioneering work of Wise *et. al* [3], researches in neuroscience and neural engineering have experienced rapid growth, especially in exploring new probe materials as well as new fabrication technologies to produce miniaturized, customized, and high-density electrode arrays for the stimulation of neurons. Despite their great potential, the electrode arrays employed in most of the current brain stimulation technologies are constantly affected by the migration of cells (such as astrocytes) around the devices, which leads to increased impedance and alterations of the electric field in the stimulation processes. One way to avoid the influence of surrounding neuron cells on the stimulation signal is magnetic stimulation, where a magnetic field is generated and is not affected by the encapsulation of astrocytes or any other cells. Transcranial Magnetic Stimulation (TMS) is a commonly used non-invasive brain stimulation technique which utilizes a strong alternating magnetic field (1.5 T to 3 T) to modulate the neuron activities [4-6]. However, due to the bulky and non-invasive nature of this setup, it is impossible to generate a highly focused magnetic field. Moreover, as the magnetic field decays exponentially over distance, this technique cannot stimulate neurons located deep inside the brain. As a complementation of TMS, Deep Brain Stimulation (DBS) implants electrodes in certain regions of the brain permanently to activate deeply located neurons [7-9]. Nevertheless, heating effects and large power consumption due to constant application of relatively large-amplitude current are major drawbacks of DBS. Consequently, the development of an implantable magnetic neurostimulator with the ability of generating highly localized magnetic field through a low power input is essential for both the study of neuron activities and the treatment of neuron disorders.

One potential nanostructure capable of magnetic stimulation of the neuron is the magnetic nanowire (NW). The displacement of magnetic domain walls through the spin transfer torque (STT) of electrons has been widely studied to switch local magnetization in high-density magnetic recording applications [10-12]. The velocity of the domain walls can be controlled by the applied current density. Pulses of highly spin-polarized current can move the entire pattern of DWs coherently along the nanowire. A neuron cell placed at a certain location on top of the NW will be stimulated with an electromotive force (EMF) which is induced by the change of magnetic stray field generated by the domain wall movement within the NW. Since the width of the domain wall is in nanometer scale, the stimulation is highly localized and is free from the influence of the surrounding environment, which facilitates neuron stimulation at the cellular level. In this paper, we have theoretically demonstrated the feasibility of stimulating an individual neuron under adjustable magnetic field strength and frequency with nano-fabricated magnetic NW arrays.

2. Methods

The magnetic dynamics including spin transfer torque (STT) terms with an extended LLG equation is used in the simulation, which can be expressed as:

$$\frac{d\mathbf{m}}{dt} = \gamma_0 \mathbf{H}_{eff} \times \mathbf{m} + \alpha \mathbf{m} \times \frac{d\mathbf{m}}{dt} - (\mathbf{u} \cdot \nabla)\mathbf{m} + \beta \mathbf{m} \times (\mathbf{u} \cdot \nabla)\mathbf{m} \quad (1)$$

$$\mathbf{u} = \frac{J_c P g \mu_B}{2 e M_s} \quad (2)$$

where $\mathbf{m} = \frac{\mathbf{M}}{M_s}$ is the unit magnetization vector, $M_s$ is the saturation magnetization, $\mathbf{H}_{eff}$ is the effective magnetic field, $\gamma_0$ is the absolute value of gyromagnetic ratio, and $\alpha$ is the Gilbert damping parameter. The last two terms on the right side of equation (1) are adiabatic and nonadiabatic torque terms, respectively. The dimensionless quantity $\beta$ represents the degree of non-adiabaticity, $\mathbf{u}$ (in $m/s$) is the effective drift velocity of the conduction electron spins, $J_c$ is the charge current density, $P$ is the spin polarization of the current, $g$ is the Landé factor, $\mu_B$ is Bohr magneton, and $e$ is the electron charge. Here, we consider permalloy ($Ni_{80}Fe_{20}$) NW with the current applied along the wire axis, and the material parameters are assumed as follows: $\alpha = 0.02$, $\beta = 0.04$, $M_s = 8 \times 10^5 \ A/m$, $P = 0.6$, and exchange constant $A = 20 \ pJ/m$ which gives an exchange length of 5 nm. The permalloy NW has dimensions of 10 μm × 8 nm × 8 nm and the simulation cell size is set to be 2 nm × 2 nm × 2 nm (Table 1). We used the object oriented micromagnetic framework (OOMMF) [13] code for simulations that solves the LLG equation incorporating the STT terms.

Table 1. Simulation Parameters of Magnetic Domain Wall Movement in Permalloy NW

| Parameter | Description | Value |
| --- | --- | --- |
| NW Dimensions [a] | Length × Width × Thickness | 10 μm × 8 nm × 8 nm |
| Cell size | Length × Width × Thickness | 2 nm × 2 nm × 2 nm |
| $\alpha$ | Gilbert damping factor | 0.02 |
| $\beta$ | Nonadiabatic spin transfer torque factor | 0.04 |
| A | Exchange constant | $20 \times 10^{-12} \ J/m$ |
| P | Polarization factor | 0.6 |
| $M_s$ | Saturation magnetization | $0.8 \times 10^6 \ A/m$ |
| $J_C$ | Charge current density | $10^{11} - 2.4 \times 10^{13} \ A/m^2$ |
| I | Charge current | 6.4 μA – 1.5 mA |

Based on the Maxwell-Faraday law, alternating magnetic flux density can induce an electromotive force (EMF):

$$\oint \mathbf{E} \cdot d\mathbf{l} = - \iint \frac{\partial \mathbf{B}}{\partial t} \cdot d\mathbf{S} \quad (3)$$

where $B$ is the magnetic flux density, $E$ is the electric field, $l$ is the contour and $S$ is the surface area. It is reported that the neuron cells can be stimulated by an electric field higher than $10\ mV/mm$ with a duration longer than $50\ \mu s$ [14-17] This can be determined by the domain wall velocity and can thus be controlled by the applied current density.

## 3. Results and Discussion

*3.1. Generation of highly localized magnetic field*

As shown in Figure 1a, due to large shape anisotropy, the magnetizations in the domains of a magnetic NW with very high aspect ratio (10 μm : 8 nm) tend to lie along the long axis of the NW (x axis), resulting in negligible out-of-plane (y or z axis) components of the stray field. To minimize the total energy, there are usually multiple domains in the NW, which are separated by the domain walls. As the magnetizations rotate either towards y or z direction within the domain wall, the out-of-plane components of the stray field become nonzero. During the current-driven domain wall motion, the out-of-plane stray field experienced by a neuron cell located at a fixed location in proximity to the NW surface will either increase or decrease, generating an electromotive force, i.e., a stimulation signal. Note that without further notation, the stray field in the following content refer to the stray field perpendicular to the NW surface that is used to generate the stimulation signal.

To realize cellular level neuron stimulation, the out-of-plane stray field should be highly localized around the domain wall, which is determined by the fast decay of the stray field and the width of the domain wall. The amplitude of the stray field along z direction ($H_z$) is plotted against the distance from the NW surface in Figure 1b. The maximum stray field at the surface of the NW is $3.2 \times 10^5\ A/m$, which is in the same order of the saturation magnetization ($8 \times 10^5\ A/m$) and decays to less than $1\ A/m$ at a distance of 150 nm. Since the size of the neuron cells are in micrometer range, while the thickness of the cell membranes is in the order of several nanometers, the distribution of the NW stray field can penetrate the membrane of the cell on top of its surface without further influence other surrounding cells, which makes it possible to realize single-cell stimulation. The distribution of the stray field in transverse direction is determined by the domain wall width. The width of the domain wall is around 40 nm for a permalloy NW with a thickness of 8 nm as observed in our simulation, which is consistent with the previously reported value and can be adjusted by simply altering the geometry of the NW [18]. This facilitates the highly localized stimulation at a specific spot of the neuron cells. As shown in Figure 1c, $H_z$ is nonzero within each domain wall in the NW and decays to zero at the edges of the domain wall. The amplitude of $H_z$ is different at each domain wall due to the difference in the magnetization orientation within the wall, which will be discussed in section 3.2. Due to the nanoscale domain structure of the permalloy NW and the fast decay of the magnetic field in non-magnetic space, highly localized stimulation signal can be generated, whose viable stimulation region can be modified by multiple parameters, including the exchange constant and the anisotropy constant of the NW material as well as the geometry of the NWs.

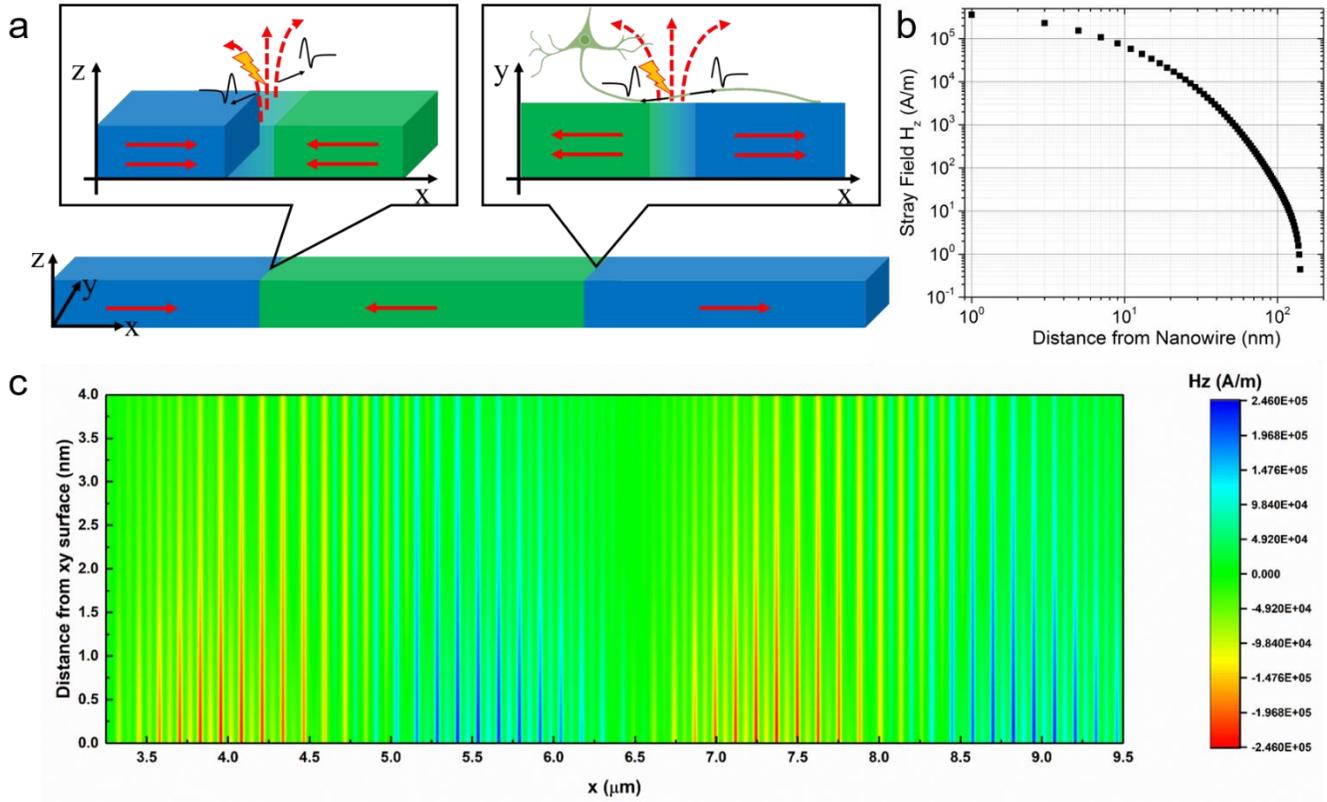

Figure 1. (a) Schematic view of different magnetic domain walls in a NW: i) perpendicular transverse wall and ii) transverse wall. Red arrow represents the magnetization in that domain. (b) Out-of-plane magnetic stray field component $H_z$ attenuates from 320 kA/m at the surface of NW to below 1 A/m at a distance of 150 nm above NW. (c) The z component of the stray field $H_z$ across the NW as a function of the distance from the xy surface of the NW.

*3.2. Current-driven domain wall motion*

To induce a change in the stray field, a spin polarized current is applied to the NW. Driven by STT, the magnetic domain wall moves along the +x direction, which can be characterized by the change in the distribution of the demagnetization field profile along the NW at different timepoints. The system is stabilized for 25 ns after the application of the current (see Figure S1 in the Supplementary Material). As shown in Figure 2a, the demagnetization field distributions are plotted at different locations along the long axis of the NW every $0.15\ ns$. The applied current $J$ is $1.2 \times 10^{13}\ A/m^2$, which corresponds to an effective drift velocity $u$ of $600\ m/s$. Each peak in the profile represents one domain wall, and all the domain walls move towards +x direction consistently after the application of the spin polarized current, which can also be confirmed by the magnetization distribution presented in Figure 2b. The domain wall profiles for $u = 400\ m/s$ and $u = 200\ m/s$, which correspondes to current densities of $8 \times 10^{12}\ A/m^2$ and $4 \times 10^{12}\ A/m^2$, also exhibit similar trend, as shown in Figure S2 in the Supplementary Material. By plotting the locations of one domain wall in the NW at different timepoints, the domain wall velocities $v$ under different applied currents can be calculated, which turn out to be $200\ m/s$, $400\ m/s$, and $590\ m/s$ under drive current densities of $4 \times 10^{12}\ A/m^2$, $8 \times 10^{12}\ A/m^2$, and $1.2 \times 10^{13}\ A/m^2$, respectively (Figure 2c). It is observed that $v \approx u$ for the utilized current densities, which is accordance with the results from other literatures when the current is above the threshold current, also known as Walker limit [19-21].

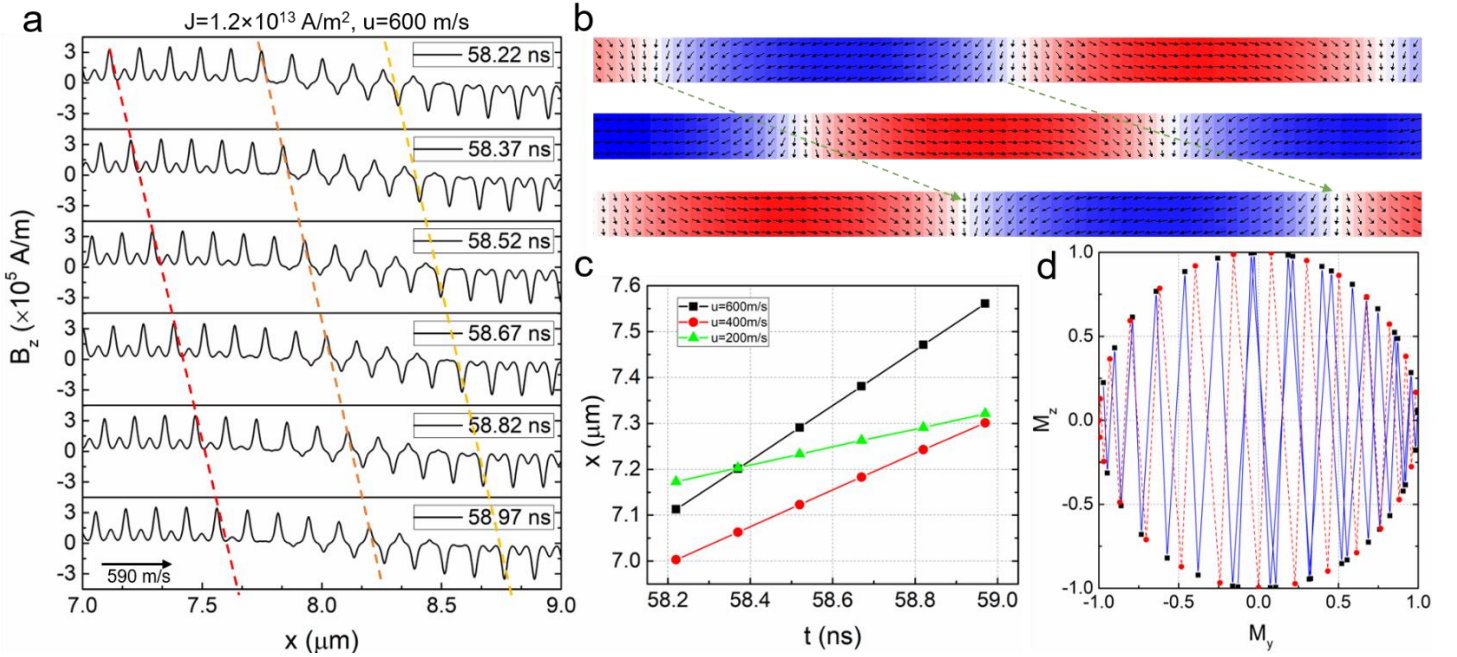

Figure 2. (a) Distribution of the demagnetization field along the center line of the NW ($y = 4nm$, $z = 4nm$) at $t = 58.22\ ns, 58.37\ ns, 58.52\ ns, 58.67\ ns, 58.82\ ns$, and $58.97\ ns$ under an applied current of $1.2 \times 10^{13}\ A/m^2$. The red, orange and yellow lines connect the locations of one domain wall at different timepoints, indicating linear displacements of the domain wall. The domain wall velocity is calculated to be $590\ m/s$. (b) Magnetization distribution in the NW at $t = 58.22\ ns, 58.37\ ns$ and $58.52\ ns$ under an applied current of $1.2 \times 10^{13}\ A/m^2$. The green arrow indicates the movement of the domain walls. (c) Domain wall position at different timepoints for $u = 200\ m/s$, $400\ m/s$, and $600\ m/s$. (d) Magnetization in y and z direction at the cross section of each domain wall along x axis under an applied current density of $1.2 \times 10^{13}\ A/m^2$. The blue line indicates the oscillation of the domain wall magnetization for domain walls located at $x = 4.019\ \mu m \sim 4.845\ \mu m$. The red line indicates the oscillation of the domain wall magnetization for domain walls located at $x = 2.565\ \mu m \sim 3.201\ \mu m$.

For transverse domain walls, when the applied current is below the Walker limit, the spin transfer torque is counteracted by an internal torque, which cants the spin out of the plane [22]. Once the current is above the Walker limit, the spin transfer torque will be much larger than the internal torque, driving the domain walls to move continuously while precessing around the x axis. The rotation of domain wall planes can be confirmed by the inconsistent peak values of $B_y$ in Figure 2(a). As shown in Figure 2(d), the magnetizations at the center of the domain wall oscillate around z axis. Since the domain walls move at a constant velocity, this effect can also be viewed as the oscillation of the domain wall planes for a given location along the NW at different timepoints, which can also be observed in the projection of magnetizations in $M_y - M_z$ plane in Figure 3(a). The reason behind this oscillation pattern can be explained qualitatively as follows. Due to dipolar interaction, the magnetizations at the end of the NW tends to tilt away from x axis. According to the first term of the LLG equation, $\gamma_0 \boldsymbol{H_{eff}} \times \boldsymbol{m}$ where $\boldsymbol{H_{eff}}$ mainly consists of the demagnetization field, the magnetizations at the end of the NW precesse around the x axis. When the damping term is taken into consideration, the magnetizations will be pushed toward x axis with a decreasing precession angle. The spin transfer torque is either in the same or opposite direction of the damping term depending on the direction of the applied current, which will either move the magnetizations toward or against x axis [23]. Upon the application of the spin polarized current, the non-zero y and z components of the magnetizations can be transferred into the NW. Consequently, the end surface of the NW can be viewed as a domain wall generator. Since the domain walls are constantly injected into the NW by the current, the equilibrium domain wall structure will depend on the possible configurations of adjacent domain walls upon collisions. According to Kunz [24], two domain walls can either annihilate with each other or form a stabilized structure depending on the types of topological defects. It was shown that only domain walls with the same topological charges can be preserved during the collision, resulting in a 360° domain wall, while the domain walls with opposite topological charges will annihilate, forming a single domain. In our cases, only adjacent domain walls with opposite signs of $m_z$ can survive, which explains the oscillation of domain wall planes around the z axis. As a result of the domain wall rotation, the demagnetization field generated by the domain wall along y and z direction exhibits a wave-like form, as shown in Figure 3 (b) and (c), where the frequency of the oscillation is determined by the distance between adjacent domain walls and the velocity of the domain walls.

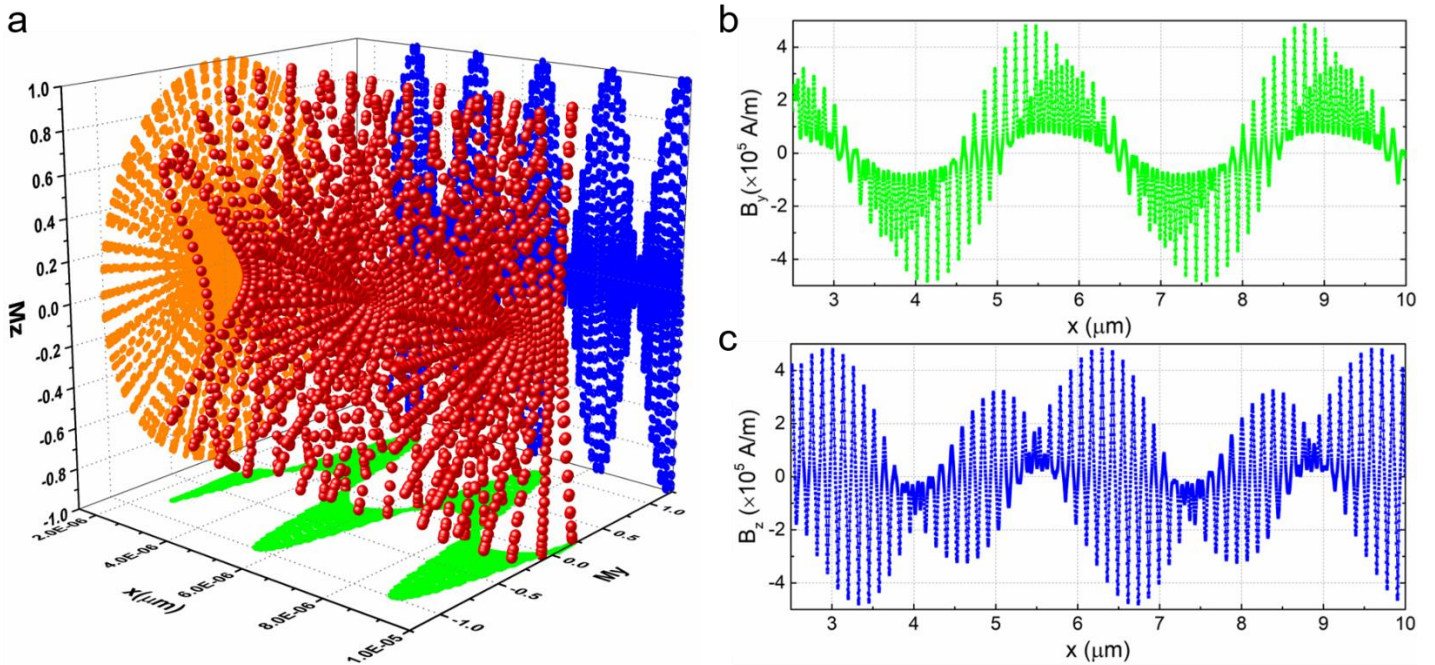

Figure 3. (a) Three-dimensional plot of the magnetizations at the center of the NW under an applied current of $1.2 \times 10^{13}\ A/m^2$. The projection in the $x - M_y$, $x - M_z$, and $M_y - M_z$ planes are also shown. The distributions of demagnetization field $B_y$ (b) and $B_z$ (c) along the x axis are calculated from (a).

*3.3. Electromotive force for neuron stimulation*

According to Faraday's Law, the alternating stray fields from the NW can generate electromotive forces. Assuming the contour of $E$ field is a circle with a diameter comparable to the width of the domain wall, the Faraday's Law can be rewritten as:

$$2\pi r E = -\frac{\Delta B}{\Delta t}\pi r^2 \quad (4)$$

Which gives

$$E = -\frac{r}{2}\frac{\Delta B}{\Delta t} \quad (5)$$

With $v = \frac{r}{\Delta t}$,

$$E = -\frac{v}{2}\Delta B \quad (6)$$

where $E$ is the electrical field, $B$ is the stray field from the NW, $r$ is half of the width of the domain wall, $\Delta t$ is the time difference between two adjacent datapoints, and $v$ is the velocity of the domain wall motion. The instantaneous electrical field is proportional to the change of the stray field and the domain wall velocity, and thus can be adjusted by tuning the magnetic properties of the NW material, or more conveniently, the applied current density. The calculated electromotive force at the surface of a permalloy NW under an applied current of $1.2 \times 10^{13}\ A/m^2$ is shown in Figure 4. Since the domain walls are moving at a constant velocity, the spatial distribution of the electrical field can be converted to the time domain. Like the patterns of the stray field, both electromotive forces along y and z axis exhibit wave-like behaviors with a maximum electrical field of $14\ V/m$, which is larger than the minimum requirement for neuron modulation ($5\ V/m$) [15]. It is worth noting that the threshold electrical field may vary with the pulse width of the stimulation signal [25]. Here, the frequency of the stimulation signal is 4.76 GHz. When the applied current densities are decreased to $8 \times 10^{12}\ A/m^2$ and $4 \times 10^{12}\ A/m^2$, the amplitudes of the stimulation signal are decreased to $7.8\ V/m$ and $3.8\ V/m$, respectively, and the frequencies are decreased to $2.86\ GHz$ and $2.78\ GHz$, respectively. The magnitude of the stimulation signal needs to be adjusted according to different clinical applications by changing the applied current density, the geometry of the NW, and/or the magnetic properties of the NW material. Since the domain wall motion can be continuously driven by the spin polarized current, the duration of the signal can be easily controlled by turning the current on and off. Based on the application, an alternating current with high frequency can also be employed to obtain the required pattern of the stimulation signal. The commonly used neuron stimulation technologies such as electrical current-based DBS and micro-coil requires electrical currents in mA range, which is not only power consuming but can also lead to heating effects. NW-based spintronic neuron stimulation, however, only requires current densities of $10^{12} \sim 10^{13}\ A/m^2$, which corresponds to currents in $\mu A$ range.

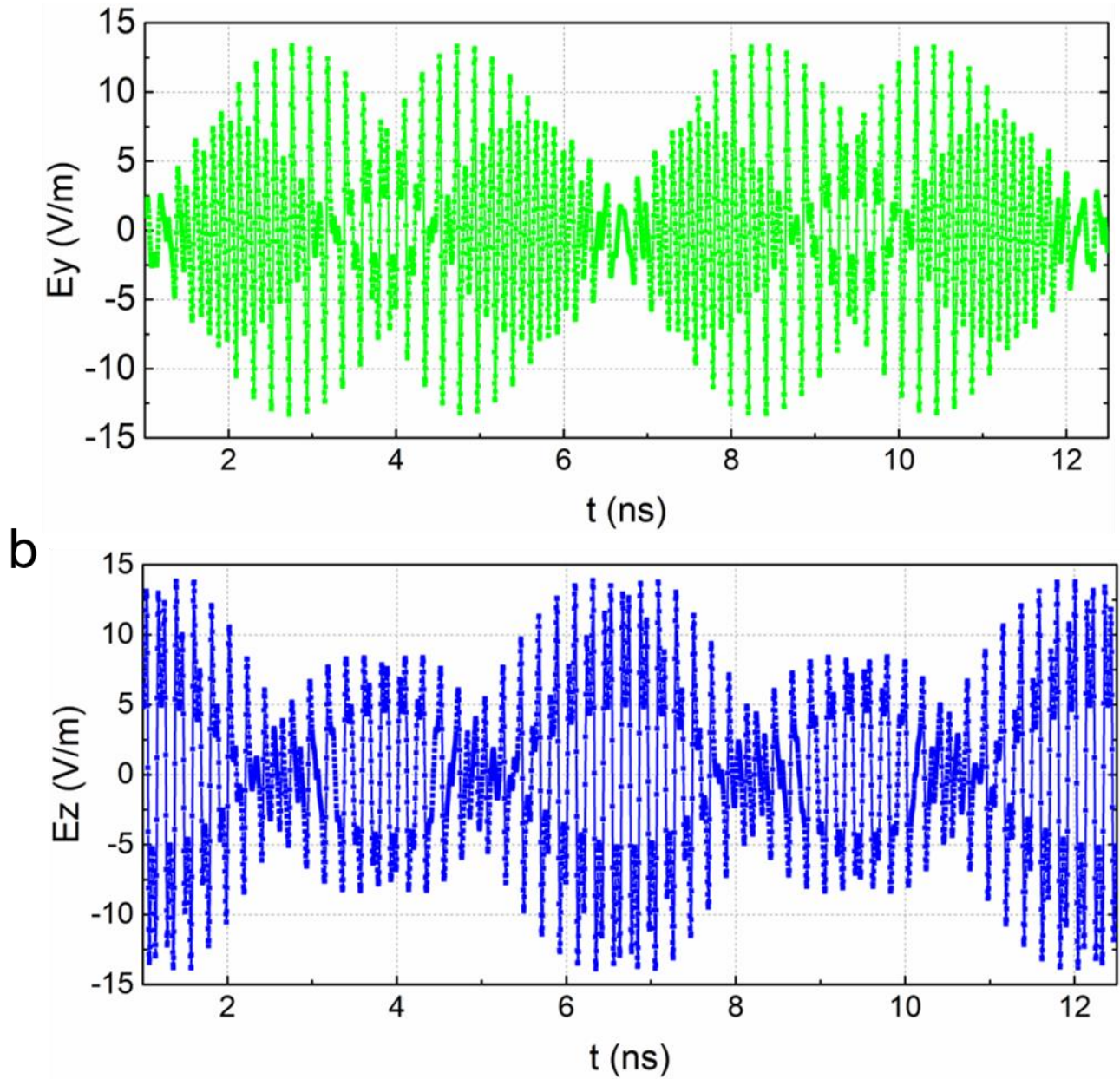

Figure 4. The electrical field generated by the alternating NW stray field during domain wall motion along (a) y axis and (b) z axis under an applied current density of $1.2 \times 10^{13}\ A/m^2$.

## 4. Conclusions

Current neuron modulation technologies suffer from multiple problems such as encapsulation of cells around the devices, bulky equipment, large power consumption and heating effects. Magnetic NW is proposed in this letter as a potential candidate with the capability of overcoming these difficulties. When the applied electrical current is above the Walker limit, the magnetic domain walls within the NW can move at a constant velocity comparable to the drift velocity of the conduction electron spins. Due to domain wall rotation, the stray fields generated by the domain walls exhibit wave-like patterns along the NW. The resulting electromotive force from the domain wall motion has a maximum amplitude of $14\,V/m$ with a frequency of 4.76 GHz. With the application of NW based spintronic nanodevices, the required electrical current can be reduced from tens of $mA$ to several $\mu A$, which significantly reduces the power consumption as well as the heating effects. Based on the specific requirements of clinical applications, the frequency and amplitude of the stimulation signal can be adjusted by altering the frequency and amplitude of the applied current, the geometry of the NW, and the magnetic properties of the NW material. Since the NWs can be fabricated with microfabrication techniques, they can be easily integrated with other functional devices such as nanosensors on flexible substrates to realize stimulation and detection on a signal chip.

**Acknowledgements**

This study was financially supported by the Institute of Engineering in Medicine of the University of Minnesota through FY18 IEM Seed Grant Funding Program, the National Science Foundation MRSEC facility program, the Centennial Chair Professorship and Robert F Hartmann Endowed Chair from the University of Minnesota.

**Conflict of Interest**

The authors declare no conflict of interest.

# Tunable Magnetic Domain Walls for Cellular-Level Therapeutic Neuromodulation: Stimulating Neurons Through Magnetic Nanowires

Diqing Su[1, #], Kai Wu[2, #], Renata Saha[2], and Jian-Ping Wang[2, *]

[1]Department of Chemical Engineering and Material Science, University of Minnesota, Minneapolis, Minnesota 55455, USA

[2]Department of Electrical and Computer Engineering, University of Minnesota, Minneapolis, Minnesota 55455, USA

[#]These authors contributed equally to this work

*Corresponding author E-mail: jpwang@umn.edu

### S1. System Stabilization

In the OOMMF simulation, the initial states of the magnetizations are randomly distributed across the nanowire. As shown in Figure S1, the total magnetization along x fluctuates significantly until it stabilizes after 25 ns.

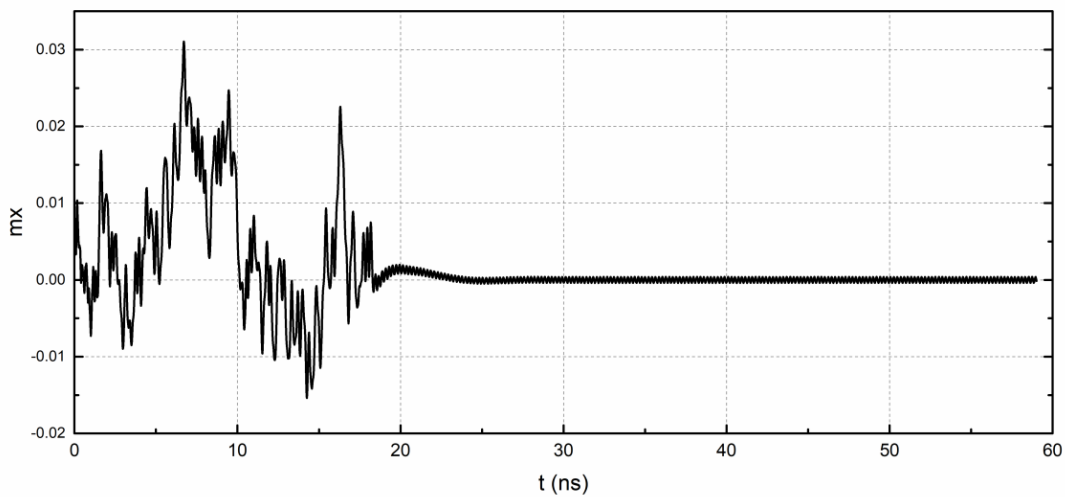

Figure S1. Alternation of magnetization along x axis after the application of a spin polarized current of $1.2 \times 10^{13}\ A/m^2$. The magnetization shown here is the total magnetization across the nanowire.

## S2. Domain wall velocity

Domain wall movement for lower current densities are shown below. There are fewer domain walls along the nanowire if the current density is decreased. The domain wall velocities under circumstances are approximately equal to the corresponding drift velocities of the conduction electron spins.

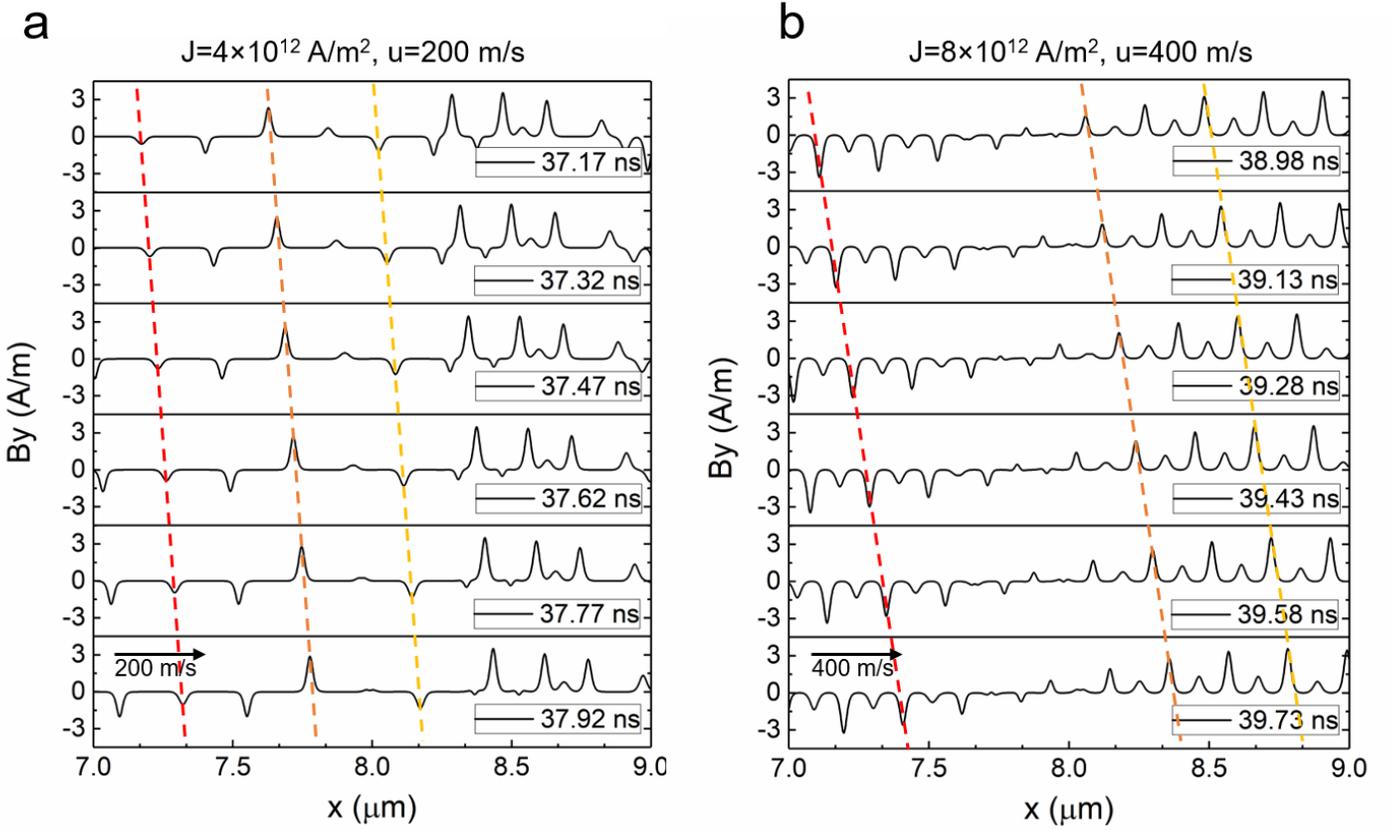

**Figure S2.** Distribution of the demagnetization field along the center line of the nanowire ($y = 4nm$, $z = 4nm$) under applied currents of (a) $4 \times 10^{12}\ A/m^2$ and (b) $8 \times 10^{12}\ A/m^2$. The time difference between each plot is 0.15 ns.

## S3. Electromotive Force

The electromotive forces generated by the nanowires for lower current densities are shown below. The maximum stimulation signal for $8 \times 10^{12}\ A/m^2$ is around 7.8 V/m with a frequency of 2.86 GHz. The maximum stimulation signal for $4 \times 10^{12}\ A/m^2$ is around 3.8 V/m with a frequency of 2.78 GHz. Consequently, both the amplitude and frequency will increase with the increase of the applied current density. The amplitude of the stimulation signal is proportional to the domain wall velocity considering the stray field generated by the domain wall is constant under different current densities while the frequency is proportional to the separation distance of two adjacent domain walls, which also increases as the domain wall velocity increases.

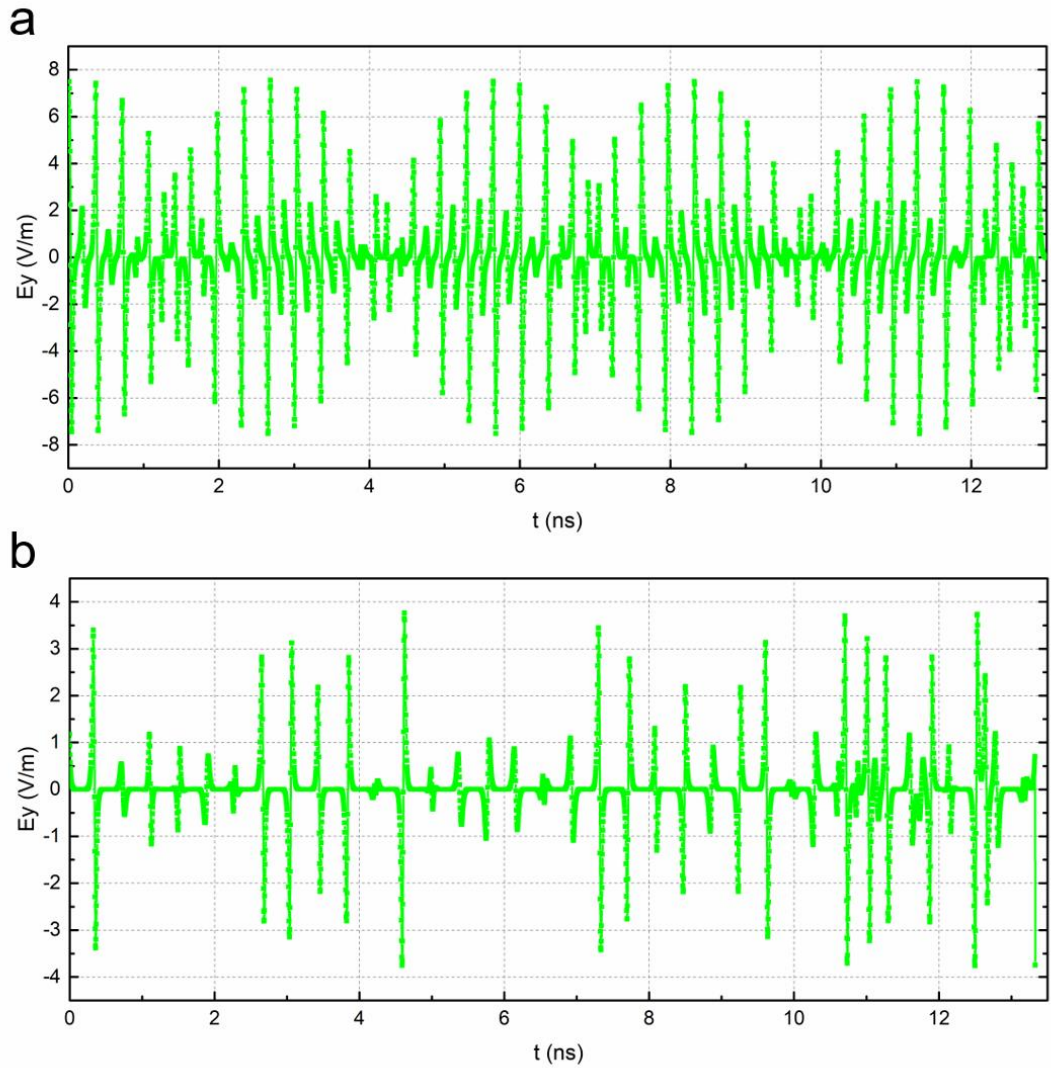

**Figure S3.** The electrical field generated by the alternating nanowire stray field during domain wall motion under applied current densities of (a) $4 \times 10^{12}\ A/m^2$ and (b) $8 \times 10^{12}\ A/m^2$.